\documentclass[sigconf]{acmart}

\renewcommand\footnotetextcopyrightpermission[1]{}
\usepackage{tabularx}
\usepackage{booktabs}
\usepackage{enumitem}
\usepackage{xspace}
\usepackage{colortbl}
\usepackage{multirow}
\usepackage{balance}
\usepackage{wrapfig}

\usepackage{pifont}
\newcommand{\cmark}{\ding{51}}%
\newcommand{\xmark}{\ding{55}}%

\newcommand{\sectopic}[1]{\vspace{0.2em}\par\noindent{\textit{\bfseries #1}}}

\newcommand{\ajr}{\textsc{AJR}\xspace}
\newcommand{\adc}{\textsc{ADP}\xspace}
\newcommand{\etal}{\emph{et al.}}

\author{Chetan Arora}
\affiliation{%
  \institution{Faculty of IT, Monash University}
  \city{Melbourne}
  \state{Victoria}
  \country{Australia}
}
\email{chetan.arora@monash.edu}

\author{Andreas Vogelsang}
\affiliation{%
  \institution{University of Duisburg-Essen}
  \city{Essen}
  \country{Germany}
}
\email{andreas.vogelsang@uni-due.de}

\author{Abbi Sharma}
\affiliation{%
  \institution{Faculty of IT, Monash University}
  \city{Melbourne}
  \state{Victoria}
  \country{Australia}
}
\email{abbi.sharma@monash.edu}

\acmConference[ASE 2026 NIER]{Proceedings of the 41st IEEE/ACM International Conference on Automated Software Engineering (ASE 2026) New Ideas and Emerging Results Track}{October 12--16, 2026}{Munich, Germany}
\acmYear{2026}

\begin{document}

\title{Specifying the Delegated-Autonomy Boundary: \\ Requirements Engineering for Agentic AI}

\begin{abstract}
Agentic AI systems do not just predict or recommend; they plan, maintain state, and act in external environments with varying degrees of autonomy. This changes the requirements engineering problem in a specific and under-addressed way: it introduces what we call the \emph{delegated-autonomy boundary}---the set of decisions about what may be delegated to the system, under what graduated authority, with what oversight, and how control is returned. Current practices bury these decisions inside prompts, tool schemas, and runtime policies, even though they are requirements-level commitments. This paper proposes two complementary artifacts. First, an \emph{Agency Justification Record} (\ajr) helps teams decide \emph{when} an agent is warranted over simpler alternatives. Second, an \emph{Agentic Delegation Policy} (\adc) captures \emph{what} must be specified for safe and effective development: purpose, authority, information, coordination, assurance, and evolution. Crucially, authority in the \adc is modelled as \emph{graduated}, i.e., a tiered structure. 
We illustrate the framework with two contrasting examples: a safety-critical hospital discharge coordination agent and an automated code review agent.
\end{abstract}

\keywords{Requirements Engineering, Agentic AI, AI agents, Delegated Autonomy, Software Engineering.}

\maketitle

\section{Introduction}

For most of software engineering (SE) history, specifying a system meant specifying its \emph{behaviour}: given this input, produce this output, subject to these constraints. The rise of AI techniques---machine learning, deep learning, and large language models (LLMs)--- has expanded RE to encompass data quality, model validation, fairness, and uncertainty~\cite{vogelsang2019requirements,Ahmad2023REAI,Habiba2024Mature}. Yet the underlying assumption held: the system \emph{responded}; it did not \emph{act}~\cite{hassan2025agentic}.
Agentic AI breaks this assumption entirely. An agent invokes tools, modifies external state, retains context across steps, and pursues goals through \emph{iterative, self-directed} sequences of autonomous decisions~\cite{OpenAI2025Guide,Anthropic2024Agents}. 
We scope this paper to such \emph{high-autonomy, multi-step} agentic
systems (with cross-system action) and distinguish them from single-step LLM inference embedded in a pipeline, which some practitioners also label ``agents'' but that introduces a different, narrower RE problem. Specifying an \emph{agentic} system requires deciding not only what it should produce but \emph{also what it may do}, \emph{to what resources}, \emph{under whose authority}, and \emph{when control returns to a human}. We call the specification of these decisions the \emph{delegated-autonomy boundary} and argue that it is the central missing object in current RE practice.

This gap arises for two recurring reasons. The first is \emph{unjustified agency}: a system is built as an agent, even though a prompted model, deterministic workflow, or conventional software design would suffice. The second is \emph{under-specified agency}: once an agentic architecture is chosen, the system's authority, memory scope, coordination rules, escalation conditions, and success criteria remain implicit---buried in prompts, tool schemas, or orchestration scripts. Together, these failure modes reveal that RE for agentic AI must address both \emph{when} to use agents and \emph{what} must be specified.

The urgency is real. Agents are already operating in healthcare coordination, software development, legal document processing, and enterprise automation~\cite{Wang2025Agents,Vaidhyanathan2025Architecture}. While robust frameworks for NL specification are emerging, overreliance on ad hoc prompt tinkering in high-risk operational contexts poses long-term governability challenges~\cite{Staufer2026AgentIndex,huang2025prompt,Tang2026AgenticSE}. For safety-critical or heavily regulated deployments, unstructured engineering compromises auditability and makes regulatory compliance a moving target. The window for establishing good practice is open now, during the formative period of agentic deployment, and our paper presents a vision in that direction. The \ajr and 
\adc structures were developed in 
consultation with two senior practitioners: one with experience in agentic enterprise systems, 
and one in healthcare, each with at least more than ten years of industry experience. Their feedback shaped both the criteria in Section~\ref{sec:ajr} 
and the authority-tier structure in Section~\ref{sec:adc}. 

\sectopic{Contributions.} We make the following contributions in this paper.

\begin{enumerate}[leftmargin=*,itemsep=2pt,topsep=4pt]
    \item \textbf{A new RE problem formulation.} The \emph{delegated-autonomy boundary} as a first-class RE concern, distinct from prior work on adaptive systems, goal modelling, or AI-specific RE.

    \item \textbf{The Agency Justification Record (\ajr).} A lightweight artifact that disciplines teams to justify agentic architectures against simpler alternatives, operationalising the principle of \emph{minimal justified autonomy}.

    \item \textbf{The Agentic Delegation Policy (\adc).} A structured specification artifact covering purpose, graduated authority, information and memory, coordination, assurance, and evolution.

    \item \textbf{A research agenda for RE for agentic AI.} Four open problems---notations, requirements-to-runtime compilation, requirements-driven evaluation, and accountability---framed as tractable empirical and design-science questions.
\end{enumerate}

We do not claim that delegated autonomy exhausts the RE problem for agentic AI. Stakeholder goals, usability, ethics, and regulatory compliance remain essential and orthogonal concerns. In fact, \adc is defined to complement the software requirements specification (SRS). SRS captures stakeholder goals, functional requirements, and quality attributes; the \adc specialises in authority, memory, and coordination dimensions unique to agentic delegation. Hence, our claim is narrower: agentic systems introduce a \emph{distinct} RE problem regarding the scope, graduation, and governance of delegated action that must now be addressed.

\begin{table*}[!htb]
\vspace*{-1em}
\small
\caption{\ajr record with running examples.}
\label{tab:ajr}
\vspace*{-1em}
\begin{tabularx}{1.04\textwidth}{@{}p{0.17\textwidth}p{0.46\textwidth}p{0.37\textwidth}@{}}
\toprule
\textbf{Criterion} & \textbf{Discharge coord. agent} & \textbf{Code review agent} \\
\midrule
C1: Task structure & Cross-system, exception-heavy; steps and order cannot be predefined \cmark & Largely enumerable; well-defined linting and diff parsing~\xmark \\
\midrule
C2: Unstructured context& Clinical notes, payer letters, ambiguous histories require NL reasoning~\cmark & Diffs are structured; rules are explicit; ambiguity is low \xmark\\
\midrule
C3: Action surface & Orchestrator delegates  actions across EHR, and payer systems to specialist sub-agents; each sub-agent operates in a scoped boundary \cmark & Source repository; read-only diff access 
only \xmark \\
\midrule
C4: Evaluable progress & Medication reconciliation status, clinician sign-off, discharge completeness are observable \cmark & Comment accuracy \& scope violations are observable~\cmark \\
\midrule
C5: Bounded risk & Clinician approval; no finalisation without sign-off; HIPAA audit log~\cmark & Developer check before posting; actions reversible \cmark \\
\midrule
C6: Advantage over baseline & 
Discharge delay and medication reconciliation error rate vs clinical audit baseline; threshold: $\geq$20\% reduction over 
90 patient episodes~\cmark & 
Comment turnaround vs existing pipeline >90 days; threshold but little projected gain to justify agentic overhead~\xmark \\
\midrule
\textbf{Verdict} & \textbf{Agentic approach justified} & \textbf{Rejected: C1--C3 + C6 fail; prompted model} \\
\bottomrule
\end{tabularx}
\vspace*{-1.5em}
\end{table*}

\sectopic{Running examples.} We ground the paper in two contrasting examples. 
The first is a hospital discharge coordination system: a multi-agent system designed to manage patient preparation for discharge from an acute care ward. A coordinating orchestrator decomposes the task across three specialist sub-agents: a {medication agent} that reads and reconciles data from an Electronic Health Record (EHR) and pharmacy records; a {payer and scheduling agent} that interfaces with insurance authorisation portals and outpatient booking systems; and a {discharge agent} that drafts patient-facing discharge instructions. The orchestrator detects unresolved conflicts between sub-agent outputs, triggers escalation, and hands off to clinical staff. The system operates in a regulated environment (e.g., HIPAA~\cite{hipaa1996}), involves clinicians and patients as primary stakeholders, and carries direct patient-safety consequences if any sub-agent acts beyond its sanctioned scope.
The second is an automated code review agent: a system proposed to analyse pull requests (PRs) in a software repository, apply configurable style rules, and post inline comments for developers. It operates in an internal development environment and carries no direct safety consequences. 



\section{When to Use Agents: the \ajr}
\label{sec:ajr}

Not every AI-enabled feature should be implemented as an agent. RE has long recognised the principle of \emph{minimal adequate specification}: a system should not be given more complexity or autonomy than its requirements demand~\cite{van2009requirements}. Applied to agentic AI, this becomes a gate: before committing to an agentic architecture, a team must justify why simpler alternatives are insufficient. We operationalise this gate as the \emph{Agency Justification Record} (\ajr)---completed before architecture is chosen and retained as a living design rationale document, analogous to an Architecture Decision Record~\cite{Nygard2011ADR} but focused on the decision to delegate autonomy. Practitioner guidance from both OpenAI and Anthropic independently reaches the same conclusion: start with the simplest solution and reserve agents for genuinely open-ended tasks~\cite{Anthropic2024Agents,OpenAI2025Guide}. The \ajr turns this intuition into an explicit, reviewable RE commitment.

\vspace{-0.5em}
\begin{figure*}[!t]
    \centering
    \vspace*{-0.5em}    \includegraphics[width=0.73\linewidth]{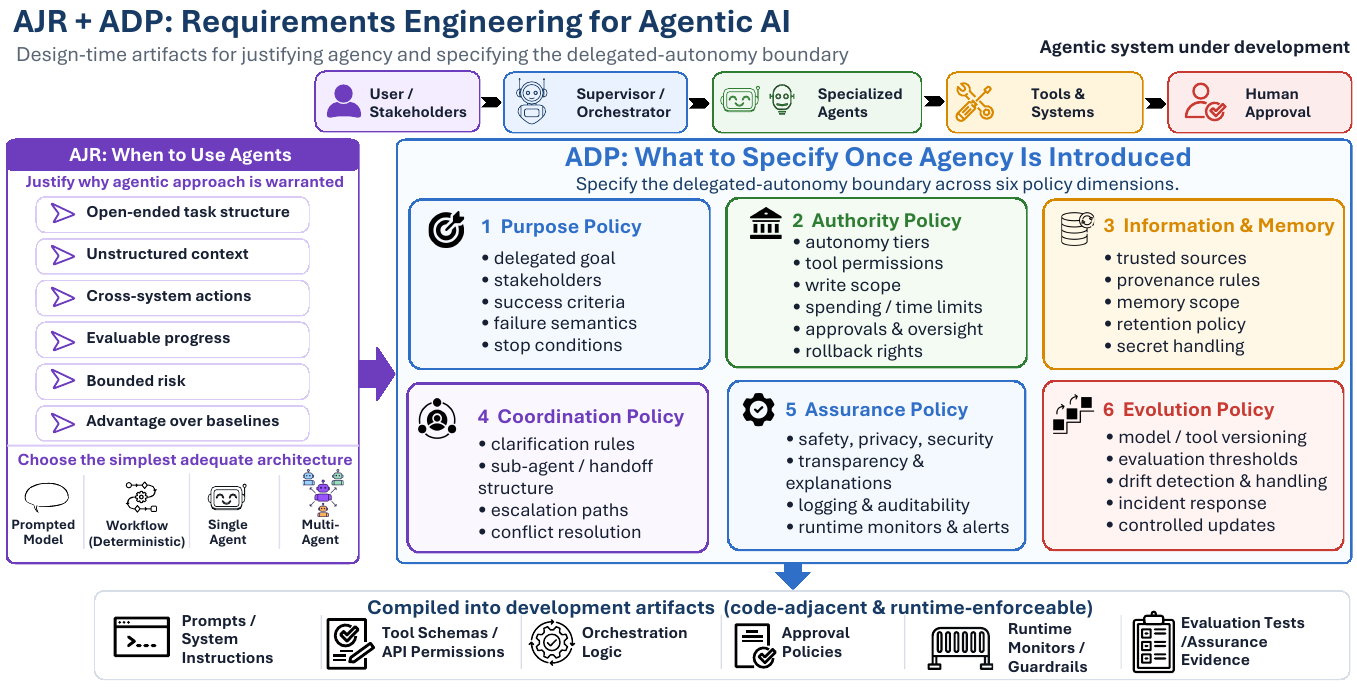}
    \vspace{-1.1em}
    \caption{RE for Agentic AI Overview}
    \vspace{-1em}
    \label{fig:framework}
\end{figure*}

\subsection{Decision criteria}
\label{sec:criteria}

A task is a strong candidate for an agent when it meets most of the following criteria. 

\begin{enumerate}[leftmargin=*,itemsep=0.8pt,topsep=1pt]
    \item \textbf{Open-ended task structure.} The number or order of steps cannot be reliably predefined; the agent must interpret context, revise plans, or recover from unforeseen exceptions.
    \item \textbf{Unstructured context.} Success depends on reasoning over NL, semi-structured artifacts, or ambiguous observations that make rule-based approaches brittle. However, a team-level decision might be necessary to allow an agentic approach if the context is deemed neither entirely ``unstructured'' nor semi-structured.
    \item \textbf{Cross-system actions.} The task must inspect, update, or orchestrate external systems rather than only generate content.
    \item \textbf{Evaluable progress.} There are explicit success criteria or intermediate checks that allow the agent or a supervising human to detect poor trajectories before harm occurs.
    \item \textbf{Bounded risk.} Harm can be constrained through sandboxes, reversible actions, rate limits, approval gates, or human takeover, and the team can specify these constraints before deployment.
    \item \textbf{Advantage over baselines.} The team can specify, in advance, a measurable threshold at which the added complexity of an agent is justified and a plan to verify. For example: \textit{``the agent must reduce discharge delays by at least 20\% over the current manual process, over 90 episodes''}. Without such a pre-committed threshold, the decision to use an agent cannot be falsified.
\end{enumerate}

Criterion~6 cannot be verified before the agent is built, but \ajr requires the team to state \emph{how} the advantage will be measured and \emph{what threshold} would justify the complexity. This forward commitment prevents post-hoc rationalisation. The \ajr must also document \emph{anti-criteria} that disqualify an agentic architecture regardless of how many positive criteria hold: a fixed workflow already performs adequately; outcomes cannot be verified; actions are irreversible or high-stakes; or multiple agents are proposed for conceptual neatness rather than demonstrated gains.

\sectopic{\ajr examples.}
Table~\ref{tab:ajr} shows \ajr records for both running examples. The discharge agent clears all six and proceeds to \adc specification; the code review scenario fails at criteria 1--3 + 6, and the \ajr rejects the agent in favour of a simpler prompted-model pipeline. The rejection path is as important as the acceptance path: a rejected \ajr records the rationale for a simpler architecture, preventing the system from accruing unjustified agentic complexity. We note that the rejection path does not mean ``no AI'': a prompted-model pipeline suffices and remains within the scope of conventional RE for AI, even though some practitioners label such pipelines ``agents'', they 
fall outside the multi-step, cross-system planning that warrants the delegated-autonomy machinery of an \adc.

\vspace*{-0.5em}
\section{What Must Be Specified: the \adc}
\label{sec:adc}
\vspace*{-0.2em}
If \ajr justifies an agent, we extend conventional RE artifacts with an \emph{Agentic Delegation Policy} (\adc). We favour \emph{``policy''} over \emph{contract} because requirements alone cannot enforce mutual obligations; a policy defines delegation terms while keeping accountability an organisational concern. The \adc is distributed across prompts, schemas, policy engines, and monitors (Figure~\ref{fig:framework} \& Table~\ref{tab:adc}). Although populated in Table~\ref{tab:adc} for illustration in second example, a rejected \ajr would not, in practice, proceed to an \adc. Crucially, \adc adheres to the `minimal adequate specification' principle by scaling with risk: safety-critical domains (e.g., healthcare) demand rigorous assurance, whereas low-risk settings (e.g., developer tooling) require only basic operational boundaries.

\vspace*{0.05em}
\sectopic{1. Purpose policy.} The specification must define the delegated goal in operational terms: intended stakeholder(s), task scope, required outputs, success criteria, and failure semantics. Critically, it must include explicit \emph{stop conditions}: max. attempts, uncertainty thresholds, deadlines, and conditions that force the system to relinquish control. In agentic systems, ``keep trying'' is itself a requirements decision, and its omission is a common source of runaway behaviour.

\vspace*{0.2em}
\sectopic{2. Authority policy and graduated autonomy.} \adc further must specify \emph{what the agent is permitted to do} and, critically, \emph{at what level of oversight}. We model it in three-tiers that align with OpenAI's guidance that agent risk depends on write access, irreversibility, permissions, and monetary impact~\cite{OpenAI2025Guide}.
\begin{enumerate}[leftmargin=*,itemsep=0.5pt,topsep=0.5pt, label=(\Alph*)]
    \item \textbf{Autonomous tier.} Actions the agent may execute without human review, e.g., read-only ops and low-cost computations.
    \item \textbf{Advisory tier.} Actions the agent may propose but not execute until a human approves, e.g., orders with discrepancies flagged.
    \item \textbf{Prohibited tier.} Actions the agent may never take regardless of instructions, e.g., irreversible high-stakes modifications, and actions outside the delegated domain.
\end{enumerate}

\vspace*{0.2em}
\sectopic{3. Information and memory policy.}
Agents mix retrieved context, intermediate notes, and persistent memory, so the teams must specify trusted sources, source precedence, retention duration, deletion rules, and confidential information handling. Session memory must be distinguished from durable memory, with requirements stating what may be remembered, for whom, and for how long. In first example, the agent is forbidden from retaining patient-identifiable information beyond the episode of care unless authorised. In second example, the agent may retain anonymised patterns to improve suggestions but not developer-identifiable commit histories.

\sectopic{4. Coordination policy.}
Requirements must define (often based on business processes) when an agent must ask clarifying questions, hand off execution, resolve contradictory outputs, or escalate to a human. 
This is vital for multi-agent designs, where decomposition introduces coordination overhead and novel failure modes~\cite{Anthropic2024Agents,Wang2025Agents}. In the discharge system, the orchestrator must reconcile conflicting outputs from the medication and payer sub-agents; the coordination policy mandates that such conflicts be escalated to a clinician rather than resolved autonomously, and forbids sub-agents from initiating out-of-scope inter-agent communication. Furthermore, the coordination policy must govern dynamic, emergent tool-chaining. Rather than restricting individual tools, it forbids conflicting tool combinations and enforces mandatory human or deterministic verification gates before downstream, state-changing actions.


\begin{table*}[!t]
\vspace*{-0.5em}
\small
\caption{Illustrative \adc specification across all six dimensions for both running examples.}
\label{tab:adc}
\vspace*{-1em}
\begin{tabularx}{1.02\textwidth}{@{}p{0.06\textwidth}p{0.04\textwidth}p{0.48\textwidth}p{0.38\textwidth}@{}}
\toprule
\textbf{Dimension} & \textbf{Tier} & \textbf{Discharge coordination agent} & \textbf{Code review agent} \\
\midrule
\multirow{2}{*}{Purpose}
  & Success  & All medication discrepancies resolved or escalated; discharge complete; clinician signed off. & Draft comments generated for $\geq$90\% of flagged lines; review turnaround reduced vs.\ baseline \\
  & Stop     & Missing chart data unresolved in $\geq$ 2h; system unavailable; clinician override. & PR closed or merged; developer dismisses all suggestions. \\
\midrule
\multirow{3}{*}{Authority}
  & Auto.    & Read patient chart; compare medication lists; retrieve scheduling. & Read diffs; run static analysis; retrieve style guide. \\
  & Advise. & Medicinal sub-agent flags discrepancy to orchestrator, which drafts consolidated discharge instructions for clinician; & Post draft inline comments pending developer acknowledgment \\
  & Prohib.  & Sign discharge orders; transmit instructions to patient; modify standing medication orders & Merge pull requests; modify protected branches; alter CI/CD configuration \\
\midrule
\multirow{2}{*}{Memory}
  & Session  & Full episode context including chart, notes; deleted at episode close & Diff context for current PR; cleared after merge or close \\
  & Durable  & Anonymised discharge-pathway patterns (no PII); authorised and logged & Anonymised code-pattern library across repositories; no developer-identifiable commit history \\
\midrule
\multirow{2}{*}{Coord.}
  & Escalate & Orchestrator escalates to clinician with unresolved conflicts after $N$ attempts; \\
  & Handoff  & Orchestrator transfers context to on-call clinician if time limit exceeded; & No sub-agent handoff; single-agent pipeline only \\
\midrule
\multirow{2}{*}{Assurance}
  & Monitor  & HIPAA audit log on every data access; escalation trigger logged with evidence bundle & All comment drafts logged with line reference and rule ID; developer action (accept/dismiss) recorded \\
  & Red-team & Prompt injection via adversarial discharge notes; privilege escalation via tool chaining & Out-of-scope branch modification; duplicate comment injection; rule-conflict resolution bypass \\
\midrule
Evolution & Govern. & Model upgrade requires regression over all tier-boundary and escalation tests; clinical sign-off & Model upgrade requires re-validation of comment accuracy and scope-boundary tests over 30-day sample \\
\bottomrule
\end{tabularx}
\vspace*{-1em}
\end{table*}

\sectopic{5. Assurance policy.}
Trustworthy-agent guidance emphasises meaningful human control, security, transparency, and privacy~\cite{Anthropic2025Trustworthy, huang2026ethical}. NIST frames these as validity, safety, security, explainability, and accountability~\cite{NIST2024Profile,NIST2023RMF}. For agentic systems, these concerns must be mapped to monitors, logs and explanation duties, including prompt-injection resilience for tool-using agents~\cite{OWASP2025PI}. 
In practice, however the assurance remains largely implicit, as 25/30 deployed agents disclose no internal safety evaluations and 23/30 have no third-party testing~\cite{Staufer2026AgentIndex}. The \adc assurance policy makes this explicit by enumerating runtime checks, red-team scenarios, and audit obligations that operationalise the assurance claims. Also, it requires that these be stated before deployment rather than after an incident.

\sectopic{6. Evolution policy.}
An agentic system is not stable if its model, tools, prompts, or policies change underneath the specification. The \adc must define versioning rules, acceptance thresholds, regression tests, rollback strategies, and incident response. Part of the requirements is specifying \emph{how the requirements remain valid over time}. In both examples, model upgrades require regression across tier boundaries and escalation conditions, not just accuracy metrics.

\section{Research Directions}\label{sec:agenda}
\sectopic{R1: Notations and tools for graduated delegated autonomy.} 
Existing goal models (i*~\cite{Yu1997istar}, KAOS~\cite{Lamsweerde2001KAOS}) and uncertainty-aware languages (RELAX~\cite{Whittle2010RELAX}) lack constructs for tiered permissions, memory scope, approvals, handoffs, and reversibility. The Agentic Problem Frames framework~\cite{Park2026APF} structures agent boundaries but omits graduated authority and memory governance. Extending these notations to represent autonomy tiers and delegation boundaries is a crucial next step; matching tool support can then be empirically evaluated to see if it improves specification quality and defect detection in controlled inspection studies.

\textbf{R2: Requirements-to-runtime compilation.} We intend to develop methods that transform \ajr/\adc artifacts into executable harnesses (tool constraints, monitors and evaluation plans). Current orchestration infrastructure~\cite{Anthropic2024Agents,OpenAI2025Guide} can directly encode some \adc components (tool permission lists, memory retention windows, approval gate triggers); security-oriented work on design patterns for prompt-injection resilience~\cite{BeurerKellner2025DesignPatterns} 
shows that some agent constraints are already partially compilable, though purpose semantics and provenance obligations require richer notation yet to be developed.

\textbf{R3: Requirements-driven evaluation and testing.} System-level testing must be guided by requirements~\cite{wang2025requirements}; accordingly, agent evaluation should be systematically derived from the \adc rather than improvised post-hoc. This formalisation directly addresses a significant transparency gap highlighted by the 2025 AI Agent Index~\cite{Staufer2026AgentIndex}, which found that most deployed frontier agents lack publicly disclosed safety evaluations. Rather than implying a complete absence of internal testing, this widespread lack of disclosure underscores an industry-wide absence of verifiable, requirements-level safety commitments. For our running examples, \adc-derived test scenarios provide this missing verifiability by targeting critical constraints: tier-boundary violations, missing provenance, and escalation pathways for medical discharge, as well as out-of-scope modifications and rule-conflict bypasses for code review. Automatically translating these \adc fields directly into concrete test oracles remains an open design-science challenge.

\textbf{R4: Human agency, trust, and accountability.} Delegating 
autonomy changes who decides, who approves, and who is responsible. 
Ronanki~\cite{Ronanki2025Trustworthy} and Holzinger 
\etal~\cite{Holzinger2025Oversight} highlight that human oversight in autonomous systems is often superficial (present in design but absent in deployed behaviour) and that accountability erodes as autonomy increases. Emerging regulatory frameworks compound this urgency~\cite{Nannini2026EULaw}. RE should make socio-technical commitments explicit: specifying not only what the agent may do, but the oversight role each stakeholder plays and the recourse available when the agent errs. Empirical work on how developers, clinicians, and auditors engage with \adc specifications, e.g., Do they update them after incidents? is needed to assess practical governability.

\sectopic{Limitations.} 
The \ajr and \adc are RE artifacts grounded in practitioner consultation and contrasting worked examples; we present them as a starting point for the community and practitioners to evaluate, refine, and build upon, with empirical validation on real-world agentic systems as the immediate next step.

\vspace*{-1em}
\section{Conclusion}
Agentic AI is not simply ``AI with tools''; it is software to which humans delegate bounded, graduated autonomy. That delegation boundary (what may be done, level of oversight, governance level) is the missing object in current RE practice. We have proposed the \ajr to discipline teams to justify agent adoption against simpler alternatives, and the \adc to specify the terms of delegation as a graduated policy. We have illustrated both artifacts with two contrasting examples spanning safety-critical regulated systems and lower-stakes development tooling. Making delegated autonomy explicit, reviewable, and testable is the central RE challenge of the agentic AI era; this paper establishes the vocabulary and artifacts that make that challenge tractable.

\section*{Data Availability Statement} There is no additional data to report for this paper.

\balance
\bibliographystyle{ACM-Reference-Format}
\bibliography{paper}

\end{document}